\documentclass[aps,pre,twocolumn,groupedaddress,showpacs,floatfix]{revtex4}
\usepackage{graphicx}
\usepackage{amsmath}
\usepackage{graphicx}
\usepackage{graphics}
\usepackage{amsfonts}
\usepackage{amssymb}

\usepackage[utf8]{inputenc}

\begin{document}
\title{Influence of diffusion on models for non-equilibrium wetting}
\author{S. R{\"o}ssner and H. Hinrichsen}
\affiliation{Fakult{\"a}t f{\"u}r Physik und Astronomie, 
         Universit{\"a}t W{\"u}rzburg, Am Hubland, 97074 W{\"u}rzburg, Germany}

\begin{abstract}
It is shown that the critical properties of a recently studied model for non-equilibrium wetting are robust if one extends the dynamic rules by single-particle diffusion on terraces of the wetting layer. Examining the behavior at the critical point and along the phase transition line, we identify a special point in the phase diagram where detailed balance of the dynamical processes is partially broken.
\end{abstract}


\maketitle
\parskip 1mm 

\section{Introduction}
In recent years there has been an considerable progress in the understanding of wetting transitions far from equilibrium. Such wetting processes are usually modelled by a non-equilibrium growth process of an interface in the vicinity of a hard-core wall which mimics an inert substrate. The dynamic rules of the growth process are chosen in such a way that detailed balance is violated, driving the system away from thermal equilibrium. Depending on the balance between growth and evaporation rates the interface may either fluctuate in the vicinity or detach from the wall, leading to a wetting transition between a pinned and a moving phase.

A simple model for non-equilibrium wetting was introduced in Ref.~\cite{NonEquWetting}. It is defined as a solid-on-solid growth process on a lattice with the restriction that the heights at neighboring sites may differ by at most one unit. While particles are deposited anywhere at the same rate, the rates for evaporation at the edges and in the middle of terraces are generally different. It was shown that such a growth process is effectively described by a Kardar-Parisi-Zhang equation~\cite{KPZ} in a potential $V$:
\begin{equation}
\label{KPZEq}
\frac{\partial{h(\vec{x},t)}}{\partial t} =
\sigma \nabla^2 h(\vec{x},t) - \frac{\partial V(h(\vec{x},t))}{\partial h(\vec{x},t)} +
\lambda [ \nabla h(\vec{x},t)]^2 + \xi(\vec{x},t)
\end{equation}
Here $\xi$ is a white Gaussian noise, $\vec{x}$ denotes the position on the substrate, $\sigma$ controls surface tension, and $V(h)$ is a potential of the form 
\begin{equation}
V(h) = \left\{
\begin{array}{cc}
\infty & \mbox{ if } h<0 \\
v_0 h  & \mbox{ if } h\geq 0   
\end{array}
\right.,
\end{equation}
where $v_0$ controls the propagation velocity of a free interface. The nonlinear term $\lambda [\nabla h(\vec{x},t)]^2$
can be shown to break detailed balance and hence drives the system away from thermal equilibrium, even if the model is in the bound phase. 

As shown in Refs.~\cite{MunozHwa,firstOrder}, the properties of wetting transitions under non-equilibrium conditions differ significantly from those at thermal equilibrium. For example, a non-equilibrium wetting transition is characterized by a different set of critical exponents. Moreover, if one introduces an additional short-range force between the substrate and the interface by adding a local well at $h=0$ in the potential $V(h)$, wetting models far from equilibrium may exhibit a phenomenon called phase coexistence, i.e., the bound and the moving phase are both thermodynamically stable in a certain region of the phase diagram.

So far these theoretical predictions have not been verified experimentally for various reasons. On the one hand it is very difficult to specify experimental conditions where detailed balance is broken on the level of microscopic transition rates. On the other hand, the models studied so far are unnatural in several respects. For example, in the model of Ref.~\cite{NonEquWetting} the transition rates are assumed to be homogeneous while realistic experiments are always affected by inhomogeneities and impurities. Another idealistic assumption of the model, on which we will focus in the present work, is the circumstance that single atoms are not allowed to diffuse although they have a finite rate for evaporation. In experiments, however, one expects that the energy barrier for single-particle diffusion is of the same order or even less than the energy barrier for evaporation. Therefore, the question arises wether the phenomena of non-equilibrium wetting observed so far are robust against diffusion of solitary particles.

Naively one may expect that the inclusion of single-particle diffusion leads to an effective term of the form $D \nabla^2 h$ in the continuum limit, which can be absorbed into the first term of the KPZ equation~(\ref{KPZEq}) without changing the critical properties. However, in what follows we assume that only single particles are allowed to diffuse, while islands of two or more atoms remain stable. In this case the dynamics is changed in non-trivial way, including interesting phenomena such as \textit{partially} broken detailed balance.

The paper is organized as follows. In Sect. 2 we review the definition of the model introduced in~\cite{NonEquWetting} and modify the dynamical rules in order to incorporate single-particle diffusion. Sect. 3 summarizes essential properties of equilibrium- and non-equilibrium wetting transitions. In Sect. 4 we first discuss a special case, where the dynamics of the model with diffusion is characterized partially broken detailed balance, while we turn to the general behavior in Sect. 5. Finally our results are summarized in Sect. 6.

\section{Definition of the model}

\begin{figure}[h]
	\begin{center}
	\includegraphics[scale=.6]{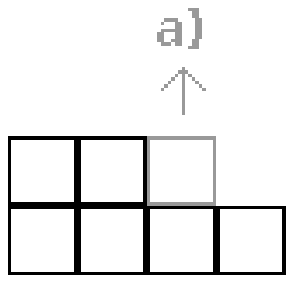}
	\includegraphics[scale=.6]{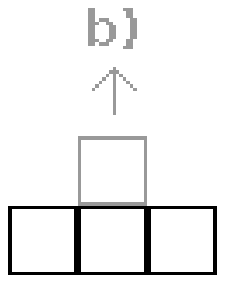}
	\includegraphics[scale=.6]{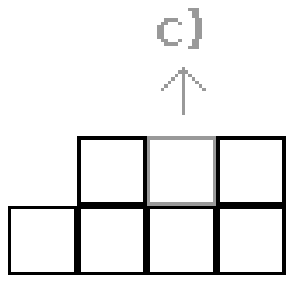}
	\includegraphics[scale=.6]{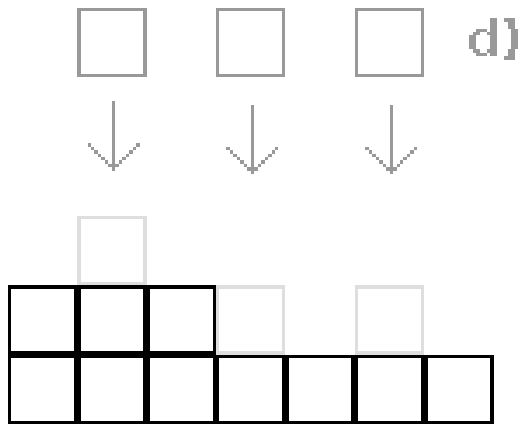}
	\includegraphics[scale=.6]{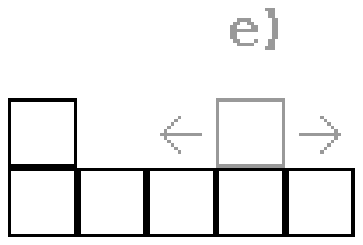}
	\caption{\label{Fig:Rules}
	Illustration of the dynamical rules. The transitions are: (a) desorption at an edge, (b) desorption of a single atom, and (c) desorption out of the middle of a plateau. Panel (d) displays three different situations where an atom can be deposited. Furthermore the additional diffusive process considered in the present work is illustrated in (e).
	}
	\end{center}
\end{figure}

The original model introduced in~\cite{NonEquWetting} is defined as follows. The form of the interface is described by height variables $h_i=0,1,\ldots$ at site $i$ of a one-dimensional lattice with $L$ sites, meaning that overhangs are not allowed. The configurations of the interface fulfill the so-called restricted solid-on-solid (RSOS) constraint, i.e., the height difference of two neighboring sites is restricted by 
\begin{equation}
\label{RSOS}
|h_i-h_{i \pm 1}| \leq 1 .
\end{equation}
The model evolves random-sequentially according to the following dynamical rules. For each update with time increment $\Delta t=\frac{1}{L}$, we randomly chose a site. Then one of the following processes is randomly selected according to specific probabilities (cf. Fig.~\ref{Fig:Rules}):
\begin{description}
\item [a)]
	An atom desorbs at the edge of a plateau with probability $\frac{r}{N}$:
	\begin{equation}
		h_i\rightarrow min(h_{i-1}, h_i, h_{i+1})
	\end{equation}
\item [b)]
	A single atom desorbs with probability $\frac{r}{N}$:
	\begin{equation}
		h_i\rightarrow h_i-1
	\end{equation}
\item [c)]
	An atom desorbs from the interior of a plateau with probability $\frac{p}{N}$:
	\begin{equation}
		h_i\rightarrow h_i-1
	\end{equation}
\item [d)] 
	An atom is deposited anywhere on the surface with probability~$\frac{q}{N}$:
	\begin{equation}
		h_i \rightarrow h_i+1
	\end{equation}
\end{description}
Here $N=\frac{1}{2r+p+q}$ denotes the normalization constant. If the moves listed above would result into a configuration that violates the RSOS constraint~(\ref{RSOS}) or produces a height less than zero the move is abandoned, time is incremented, and the next site is selected. These rules can be generalized easily to higher dimensions.
	
In this model there is no diffusion of solitary atoms along the surface. As outlined in the Introduction, one possible step to make the model more realistic is to introduce an additional rule for explicit diffusion. More specifically, a single atom is now allowed to diffuse a single step along a plateau at rate $D$ by the following rule:

\begin{description}
\item [e)] Diffusion of a single atom on a plateau to the right or to the left with equal probability $\frac{D}{N}$:\newline
	\begin{equation}
		h_i\rightarrow h_i-1\qquad h_{i\pm 1}\rightarrow h_{i\pm 1}+1
	\end{equation}
\end{description}

Again this move is carried out only if the resulting configuration fulfills the RSOS constraint with non-negative heights. Note that this additional rule is not in conflict with the previous rules listed above. The normalization has to be replaced by $N=\frac{1}{D+2r+p+q}$ so that the probabilities change accordingly.

\section{Phenomenology of the transition without diffusion}
\label{PhenomenologySection}

\begin{figure}[h]
	\begin{center}
	\includegraphics[width=85mm]{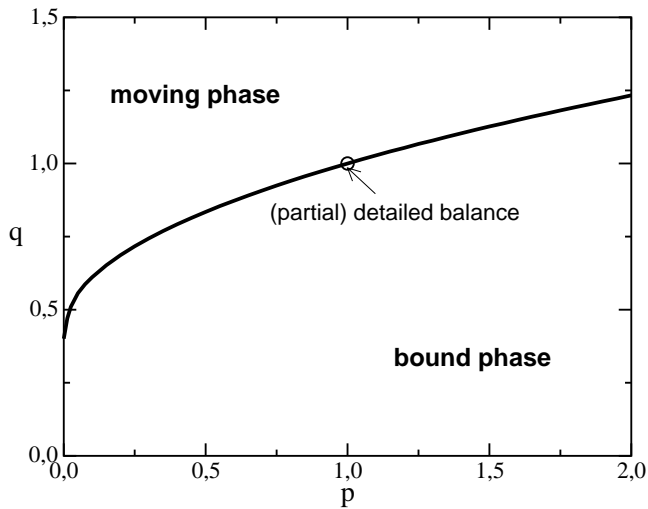}
	\caption{ \label{Fig:Phasediag}
	Phase diagram of the model for non-equilibrium wetting introduced in~\cite{NonEquWetting} in one dimension. A line of continuous phase transitions separates the pinned and the moving phase. For $p=1$ the dynamical processes obey detailed balance (see text). Introducing diffusion shifts the phase transition line slightly but the overall structure of the phase diagram remains the same.
	}
	\end{center}	
\end{figure}

Let us first review the properties of the wetting transition in the original model \textit{without} diffusion. Obviously, for a large growth rate $q$ the interface propagates while for sufficiently small $q$ it fluctuates near the wall. One can easily conclude that there is a critical growth rate $q_c$ where the propagation velocity of a free interface is zero. This defines a line of second-order phase transitions, which is shown in Fig.~\ref{Fig:Phasediag}. This unbinding transition was shown to be continuous.

Choosing $p=1$ and $q<1$ it was shown in~\cite{NonEquWetting} that the dynamical rules (a)-(d) obey detailed balance. This means that for any microscopic transition the reverse transition also exists and that the corresponding probability currents in the stationary state compensate each other. Therefore, the bound interface approaches an equilibrium state in which the probability distribution of interface configurations is given by the Boltzmann ensemble.

\begin{figure}[h]
	\begin{center}
	\includegraphics[width=85mm]{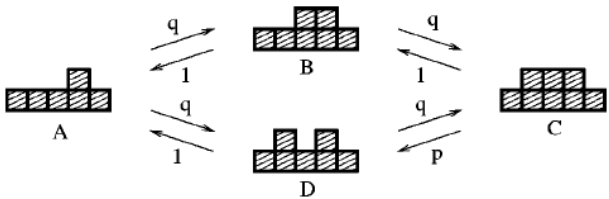}
	\caption{\label{Fig:graph_wettNonEqu}
	Breaking of detailed balance for $p \neq 1$. The figure shows four possible configurations of the interface together with the corresponding transition rates. If $P_A$ denoted the probability to find the interface in configuration $A$, detailed balance applied along the upper sequence of transitions would imply $P_C=q^2 P_A$ while the lower sequence of transitions would give $P_C=(q^2/p)P_A$. Therefore, this counterexample demonstrates that detailed balance is broken for $p \neq 1$.}
	\end{center}
\end{figure}

For $p \neq 1$, however, it can be shown that detailed balance is broken (see Fig.~\ref{Fig:graph_wettNonEqu}). As before, the critical point $q_c$ can be determined by readjusting $q$ in such a way that the asymptotic growth velocity is zero. 

A free interface is known to roughen algebraically in such a way that the width increases as
\begin{equation}
w \sim t^\beta f(t/L^z)\,,
\end{equation}
where $f$ is a scaling function, $L$ is the lateral system size, and $\beta$ and $z$ are the roughening and the dynamical exponent, respectively. It can be shown that for $p=1$, where the bound phase exhibits detailed balance, the transition belongs to the so-called Edwards-Wilkinson universality class~\cite{EW}, which is characterized by the exponents $\beta = 1/4$ and $z=2$. In the non-equilibrium case $p \neq 1$, however, the asymptotic roughening behavior belongs to the KPZ universality class with the exponents $\beta=1/3$ and $z=3/2$ in one dimension~\cite{KPZ}. 

Introducing a wall at zero height induces an unbinding transition at $q=q_c$, which may be characterized by the density of contact points $\rho_0(t)$, where the interface has zero height. At the transition this quantity decays algebraically as $\rho_0(t) \sim t^{-\theta}$, giving rise to an additional exponent~\cite{MN1,MN2,Kissinger}
\begin{equation}
\theta = \begin{cases}
1.18(1) & \mbox{ for } p<1 \\
3/4     & \mbox{ for } p=1 \\
0.23(1) & \mbox{ for } p>1 
         \end{cases}.
\end{equation}
This means that non-equilibrium wetting can be categorized into three different universality classes, namely, the Edwards-Wilkinson class for $p=1$ and two different KPZ-like classes depending on the sign of $\lambda$ in the KPZ equation, i.e., on $p-1$.

To each point of the phase transition line there is a profile $P(h,t)$ that describes the probability to find the interface at height $h$. This probability distribution scales as
\begin{equation}
\label{KPZScaling}
P(h) = t^{-\beta} \, g(h t^{-\beta})\,,
\end{equation}
where $g$ is another universal scaling function describing the form of the profile. Therefore, profiles measured at different times can be collapsed onto one curve by plotting $P(h) t^\beta$ against $h t^{-\beta}$. This method also allows one to estimate the exponent $\beta$. As a major breakthrough, Pr{\"a}hofer and Spohn were able to compute the scaling function $g$ for the case of a \textit{free} interface analytically~\cite{Spohn}. However, for non-equilibrium wetting (i.e. with a wall at zero height) the scaling function is still unknown although it could be approximated recently by mean-field techniques~\cite{Ginelli}. In the following we investigate to what extent these results are robust with respect to diffusion of single particles.

\section{Transition with diffusion for $p=1$}

Let us now turn to the generalized model for non-equilibrium wetting which includes diffusion of single particles (rule (e) in Fig.~\ref{Fig:Rules}). First we note that this process always breaks detailed balance, even for $p=1$. This happens when an atom diffuses towards the edge of an island where it sticks irreversibly. Obviously, this process cannot be reversed. So the interesting question is whether diffusion leads to a different type of critical behavior.

\begin{figure}[h]
	\begin{center}
	\includegraphics[width=85mm]{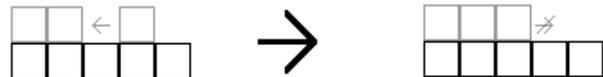}
	\caption{ \label{Fig:diff}
	Breaking of detailed balance by single-particle diffusion. A particle diffuses to the edge of an island where it sticks irreversibly. Obviously, this process cannot be reversed so that detailed balance is broken.}
	\end{center}
\end{figure}

\begin{figure}[h]
	\begin{center}
	\includegraphics[width=85mm]{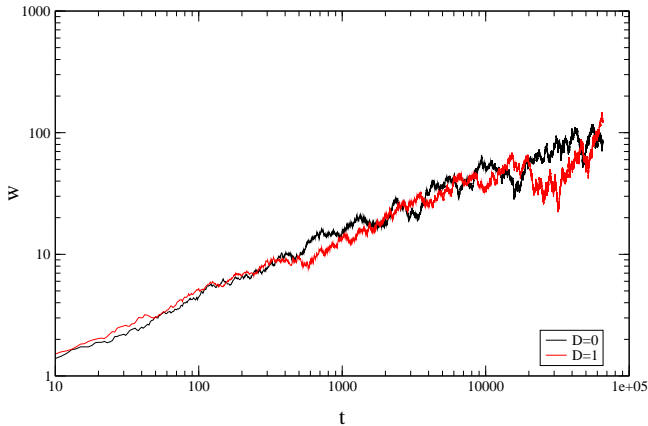}
	\caption{\label{Fig:width}
	This graph shows that the width scales according to the Edwards-Wilkinson universality class as $w \sim t^{1/4}$.
	}
	\end{center}
\end{figure}

To answer this question we performed standard Monte Carlo simulations of a 1+1-dimensional system with $L=4096$ sites and $T=65536$ time steps with and without diffusion, averaging over $100$ independent runs. As shown in Fig.~\ref{Fig:width}, the width increases as $t^{1/4}$ in both cases. Surprisingly, detailed balance can be broken without changing the critical point $q_c=1$ and the exponent $\beta=1/4$, i.e., without driving the transition away from the Edwards-Wilkinson universality class. However, as will be shown below, it is possible to understand this apparent contradiction.

\subsection{Invariance of the critical point for $p=1$}\label{SubsecA}

\begin{figure}[h]
	\begin{center}
	\includegraphics[width=85mm]{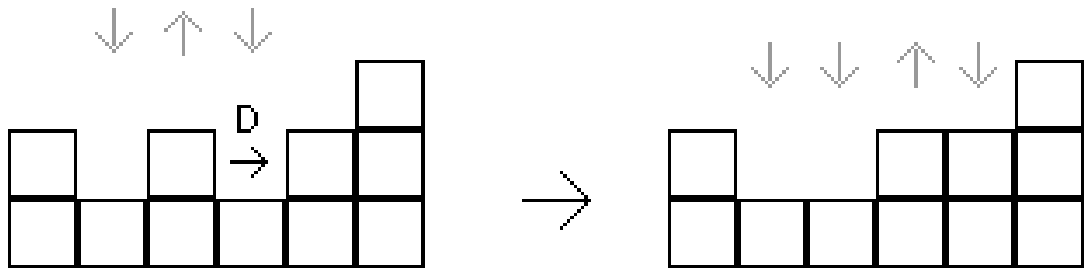}
	\caption{\label{Fig:first_conf}A possible configuration for a diffusive transition from site number $3$ to site number $4$. The arrows indicate those sites where atoms could adsorb or desorb, respectively. Obviously, the diffusive move shown here increases the number of sites susceptible to deposition by 1.}
	\includegraphics[width=85mm]{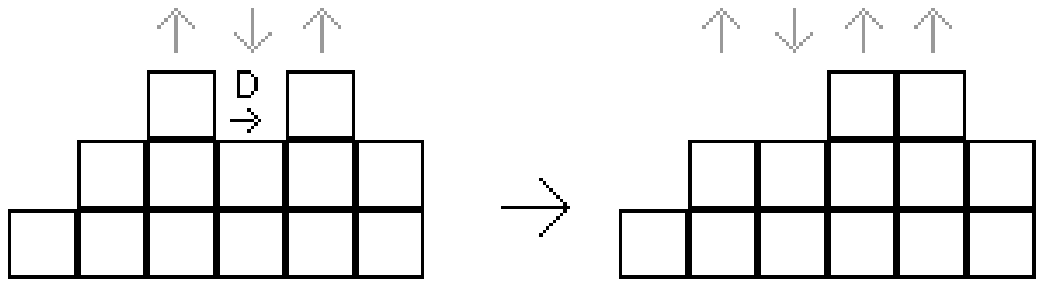}
	\caption{\label{Fig:second_conf}The mirror-symmetric configuration of the configuration shown in Fig.~\ref{Fig:first_conf} obtained by reflections $h \to -h$ and $x \to -x$. Here a diffusive move increases the number of sites susceptible to \textit{desorption} by 1, compensating the contribution of the configuration shown in Fig.~\ref{Fig:first_conf}.} 
	\end{center}
\end{figure}

Let us first examine the possible transitions for each configuration of a \textit{free} interface. For example, Fig.~\ref{Fig:first_conf} shows a particular configuration of $6$ sites where a particle diffuses from site $3$ to site $4$. Initially we set $D=0$ and $q=q_c=1$, meaning that after sufficiently long time all allowed configurations occur with the same probability. Counting all sites where atoms can be deposited and similarly all sites from where particles can desorb we quantify how this segment of the interface contributes to the average growth velocity in the next instance of time. Since the asymptotic propagation velocity at the transition is zero, these probabilities for deposition and evaporation averaged over all possible configurations have to be equal. Then, switching on diffusion, we pose the question whether diffusive moves destroy this balance or not. 

Indeed, for some configurations such as the one shown in Fig.~\ref{Fig:first_conf} the number of potential adsorption and desorption sites before and after the diffusive move changes. However, for each configuration there exists a mirror-symmetric configuration obtained by reflections $h \to -h$ and/or $x \to -x$, as demonstrated in~Fig.~\ref{Fig:second_conf}. As can be seen, this mapping turns potential absorption sites into desorption sites and vice versa. Since both of them occur with the same probability, the averaged change of the susceptibilities for adsorption and desorption cancel each other, meaning that the growth velocity remains zero even when diffusion is turned on. This argument proves that for $p=1$ the transition point $q_c=1$ is independent of the diffusion rate $D$.

\subsection{Invariance of the height profiles for $p=1$} \label{SubsecB}

Taking a closer look at the violation of the detailed balance one can see that it is only a partial violation.
First we note that diffusion does not change the total height $H=\sum_i h_i$, while all other transitions change $H$ by one unit. Therefore, the configuration space of the system can be decomposed into \textit{sectors} characterized by the value of $H$. Within those sectors diffusion is the only transition, while all other transitions take place between the sectors. Together with the previous finding that the total rates for desorption and adsorption between the sectors is invariant we conclude that the probability to find the system in a particular sector does not depend $D$. Moreover, it is obvious that diffusive moves within the sectors do not change the height profile $P(h)$. This means that the height profile does not depend on the value of $D$ and the dynamics \textit{between} the sector still obeys detailed balance. Therefore, the present model provides an interesting example of a system where detailed balance is \textit{partially} broken.

\section{Transition with diffusion for $p \neq 1$}

For $p \neq 1$ we expect that the critical point $q_c$ depends on the diffusion rate. For example, for the configurations shown in Fig.~\ref{Fig:first_conf} and Fig.~\ref{Fig:second_conf} the aforementioned compensation mechanism  is no longer possible since rates for deposition $q<1$ and desorption at the edges $r=1$ are different. In fact, considering all possible configurations of a segment of six sites, one finds that a diffusive move creates $7$ additional sites where an atom could desorb with rate $p$ and $6$ additional sites where an atom could be deposited with rate $q$ while there will be one site less from where an atom could desorb with rate $r$. As the rates in the nonequilibrium case along the phase transition line are not equal any more the phasetransition line \textit{with} diffusion is different from the one without diffusion.

\begin{figure}[h]
	\begin{center}
	\includegraphics[width=85mm]{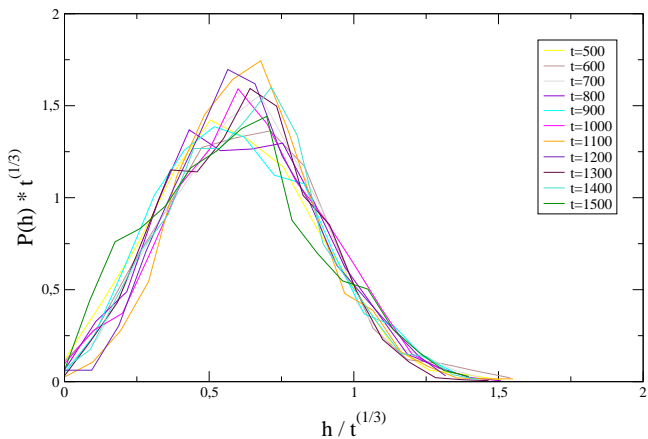}
	\caption{\label{Fig:coll_heigth}Data collapse for $p=0.3$ for different times}
	\end{center}
\end{figure}

In order to verify whether diffusion changes the critical behavior we determined interface profiles at the critical point. As mentioned in Sect.~\ref{PhenomenologySection} different curves of profiles recorded at different times can be collapsed onto a single curve by using the scaling form~(\ref{KPZScaling}) with the correct scaling exponent $\beta$. For $D=1$ we find that the critical point does not change significantly. Collapsing our profiles for with the same value of $\beta$ we obtain a convincing data collapse, supporting that for $p\neq 1$ diffusion does not change the universality class of the transition. 

\section{Behavior in the limit $D \to \infty$}

So far we discussed the behavior of the extended model for finite values of the diffusion rate $D$. How robust are the results when taking $D$ to infinity? To study this limit efficiently, we reformulate the dynamic rules without changing the definition of the model as follows. After each update we identify isolated particles that could possibly diffuse. Each of them is found in one of the following local configurations and evolves accordingly:
\begin{itemize}
\item [1)] Single particles on a plateau continue to diffuse at infinite rate (unable to leave the island) until another atom is deposited on the plateau. This diffusion process is not simulated explicitly, instead the nucleation site can be chosen randomly in agreement with the definition of the model.
\item [2)] Single particles diffusing on a terrace instantaneously attach at the neighboring island.
\item [3)] Single particles in a well will instantaneously stick randomly to one of the neighboring islands with the probabilities
\begin{equation}
p_l=\frac{d_r}{d_l+d_r} \,, \quad
p_r=\frac{d_l}{d_l+d_r}
\end{equation}
where $d_l$ and $d_r$ denotes the distances to the left and the right island, respectively. Note that this linear law holds only in the one-dimensional case.
\end{itemize}

\begin{figure}[h]
	\begin{center}
	\includegraphics[width=85mm]{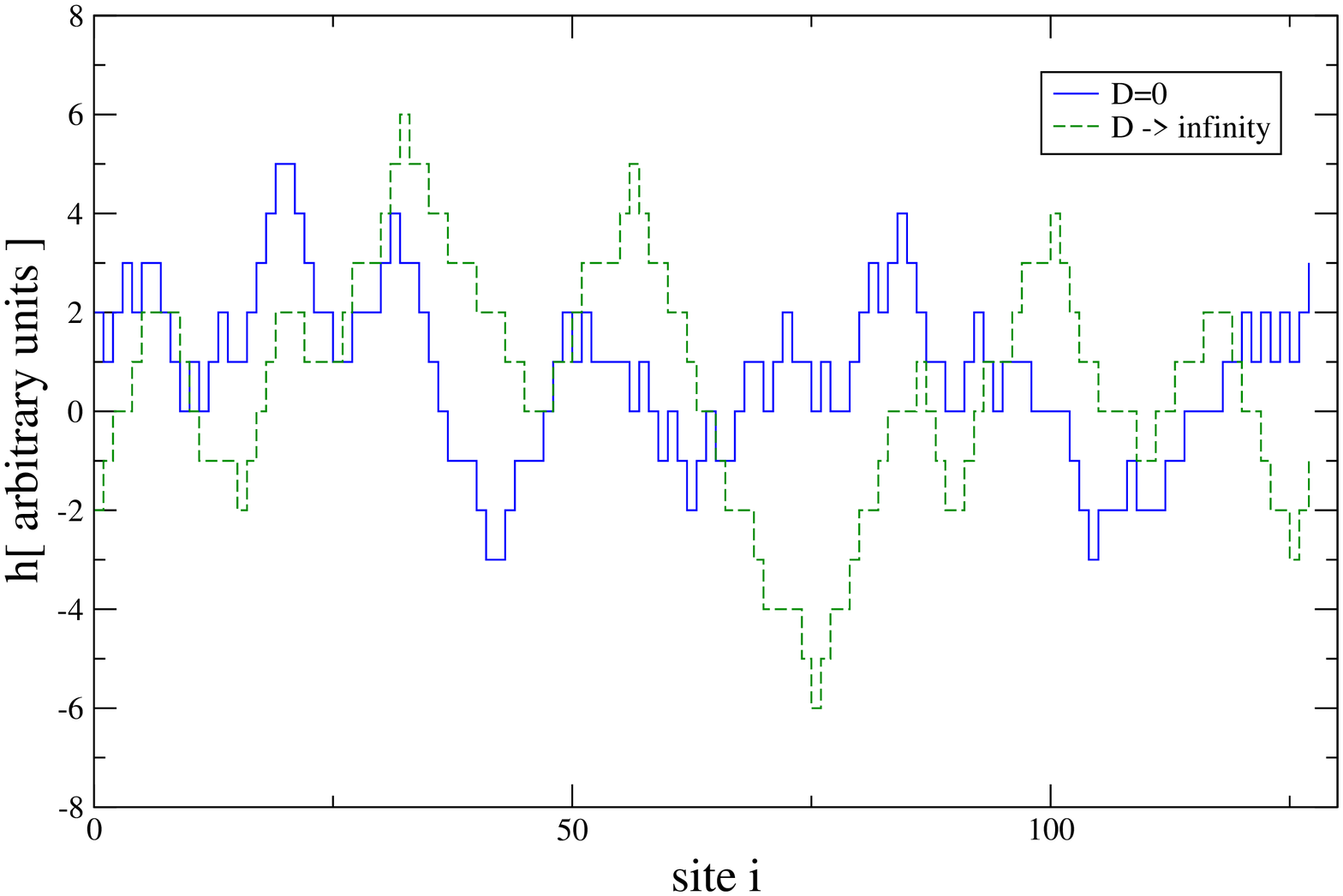}
	\caption{\label{Fig:snapshots}
	Snapshots of a free interface after 100 time steps with infinite and without diffusion. The profile with diffusion is characterized by the absence of single atoms.} 
	\end{center}
\end{figure}

Looking at typical snapshots of the interface for $D=0$ and $D=\infty$ in Fig.~\ref{Fig:snapshots} there is no big difference apart from the fact that in the latter case there are no single atoms any more. Carrying out numerical simulations similar to those presented in the previous sections we obtain similar results, in particular, the height profiles for $D=0$ and $D \to \infty$ can be collapsed onto a single curve in each of the three cases $p<1$, $p=1$, and $p>1$. This means that the critical exponents do not depend on $D$ so that the universality class of the transition is not changed by single-particle diffusion, even in the limit $D \to \infty$. This result is plausible since the arguments discussed in Sects.~\ref{SubsecA},\ref{SubsecB} do not depend on the value of $D$.

\section{Conclusions}

In this paper we have generalized a previously introduced model for non-equilibrium wetting in such a way that freshly deposited adatoms are allowed to diffuse. This generalization is necessary to compare this model with possible experimental applications, where the energy barrier for horizontal diffusion is usually of the same order or even less than the energy barrier for evaporation. As a main result we observe that single-particle diffusion in 1+1 dimensions does not change the critical behavior at the transition, i.e., all predictions made for non-equilibrium in previous works remain valid.

The results for the special point $p=1$, where the original model exhibited detailed balance in the bound phase, are particular surprising. Although  diffusion breaks detailed balance and even changes the visual appearance of the interface (see Fig.~\ref{Fig:snapshots}), the transition point $q_c=1$ as well as the rescaled height profile remain invariant, even in the limit $D \to \infty$. This apparent contradiction can be resolved by observing that detailed balance is only \textit{partially} broken within sectors of constant integrated height. This example demonstrates that broken detailed balance does not always drive the system away from equilibrium.

Extending models for non-equilibrium wetting by single-particle diffusion is only a first step towards a more realistic modelling aiming at a quantitative comparison with possible experiments. In addition, a detailed survey of energy barriers and the corresponding transition rates under typical experimental conditions would be needed. Moreover, it would be interesting to examine the influence of disorder caused by inhomogeneities and defects of the substrate.

\noindent
\textbf{Acknowledgements:}\\
We would like to thank D. Mukamel for useful discussions.


\end{document}